\documentclass{4arXiv}
\copyrightyear{2005}
\pubyear{2005}

\begin{document}
\firstpage{1}

\title[GPU-Acceptance-rejection algorithm for SSA]{An efficient GPU acceptance-rejection algorithm
  for the selection of the next reaction to occur for Stochastic Simulation Algorithms}

\author[Neri \textit{et~al}]{Neri Mickael\,$^{1}$ and Denis Mestivier\,$^{1}$\footnote{to whom correspondence should be addressed}}
\address{$^{1}$
Institut Jacques Monod, CNRS UMR 7592, Universit\'{e} Paris Diderot, Paris Cit\'{e} Sorbonne, Paris, France}

\history{\textit{arxiv.org} version - March, 2014}

\editor{}

\maketitle


\begin{abstract}

\section{Motivation:}
The Stochastic Simulation Algorithm (SSA) has largely diffused in
the field of systems biology. This approach needs many realizations
for establishing statistical results on the system under study.
It is very computationnally demanding, and with the advent
of large models this burden is increasing. Hence
parallel implementation of SSA are needed to address these needs.

At the very heart of the SSA is the selection of the next reaction to
occur at each time step, and to the best of our knowledge all
implementations are based on an inverse transformation method.
However, this method involves a random number of steps to select this
next reaction and is poorly amenable to a parallel implementation.

\section{Results:}
Here, we introduce a parallel acceptance-rejection algorithm to select
the K next reactions to occur. This algorithm uses a deterministic
number of steps, a property well suited to a parallel implementation.
It is simple and small, accurate and scalable. We propose a Graphics
Processing Unit (GPU) implementation and validate our algorithm with
simulated propensity distributions and the propensity distribution of
a large model of yeast iron metabolism. We show that our algorithm can
handle thousands of selections of next reaction to occur in parallel
on the GPU, paving the way to massive SSA.

\section{Availability:}
We present our GPU-AR algorithm that focuses on the very heart of the
SSA. We do not embed our algorithm within a full implementation in
order to stay pedagogical and allows its rapid implementation in
existing software. We hope that it will enable stochastic modelers to
implement our algorithm with the benefits of their own optimizations.

\section{Contact:} \href{mestivier@ijm.univ-paris-diderot.fr}{mestivier@ijm.univ-paris-diderot.fr}
\end{abstract}


\section{Introduction}

It is now widely acknowledged that stochasticity is an inherent
feature of many biological systems, mainly due to the small
populations of certain reactants species
(\cite{McAdams1997,Thattai2001,Paulsson2004}). This inherent
randomness cannot be dealted with deterministic approaches, and
stochastic simulations are required in order to achieve more accurate
simulations.

For complex networks, such as encountered in systems biology, each
chemical species can be implicated in many interactions with the other
components. This results in complex temporal behaviors and computer
simulations are an essential tools in order to understand its dynamics
(\cite{Achcar2011,Endy2001,Arkin1998}).

The Stochastic Simulation Algorithm (SSA) was introduced by
D.T. Gillespie more than 30 years ago
(\cite{Gillespie1976,Gillespie1977}), and has now diffuses among
different scientific communities, in particular in the field of systems
biology. Many implementations have been proposed
(see~\cite{Pahle2008} for a review,
\cite{LeNovere2001,Zhou2011,Stoll2012}) in order to popularize its use
among modelers and biologists.

However, stochastic systems need many realizations in order to capture
enough statistical information on the system under study, and are thus
computationnally demanding. With the advent of large models in systems
biology, this burden is still increasing.

As an illustration, a recent model for iron homeostasis in the yeast
\textit{Saccharomyces cerevisiae} incorporates 641 species and 1029
reactions and is simulated using a boolean version of the SSA
(\cite{Achcar2011}). Such large models have to be challenged using a
huge amount of datasets in order to assess their correctness, at the
cost of many simulations. As an example, the sensitivity analysis
involves testing 9 different values of the parameter of each of the
1029 reactions, with 1000 realizations (See: Additional File 2
in~\cite{Achcar2011}) for computing statistics, hence such a
sensitivity analysis needs $9 \times 1029 \times 1000 = 9261000$
realizations. Moreover, this model of iron homeostasis was also
validated by the confrontation of the output of the simulations mimicking
different biological contexts to more than 190 phenotypic mutants,
metabolomic data and the global analysis of more than 180 \textit{in
 silico} mutants.

In order to tackle this ``need for power'', two directions of research
have been explored over the last few years: the first accelerates
individual simulations starting from the original formulations
proposed by Gillespie~\cite{Gillespie1976,Gillespie1977}. The second
exploits parallel architectures in order to run multiple realizations
of the algorithm.

As will be developped in the Methods section, the system under study
is defined by $N$ chemical or biological species and by $M$ chemical
reactions that describe their interactions. The SSA characterizes each
reaction by a propensity function $\alpha_j$ that allows one to
compute, at each iteration of the algorithm, the probability that this
specific reaction will be the next to occur.

It is well-known that this heart of the SSA is the most
computationnally costly part of the algorithm~\cite{Komarov2012}.

Hence, the acceleration of individual simulation mainly focuses on
modifications of this selection of the index $j$ of the next reaction
to occur (see~\cite{Li2008} for a review).

For example, Gibson (\cite{Gibson2000}) introduced the Next Reaction
Method (NRM), and a dependency graph which lists the propensity
functions that depend on the outcome of each reaction, enabling them
to identify and alter only propensity functions which require
updating.
Cao (\cite{Cao2004}) proposed an Optimized Direct Method (ODM), with
a modification of the selection of the next reaction to occur that
implies the ordering of propensity functions in a search list so that
reactions occuring more frequently are the first in the list.
Another proposed acceleration by McCollum et al.
(\cite{McCollum2006})
is the Sorting Direct Method (SDM) that dynamically changes
the reaction order, which changes the propensity functions
order, the more probable being the first.
In the same spirit of reducting search depth, Li (\cite{Li2006})
introducted the Logarithmic Direct Method (LDM) which uses a binary
search to determine which reaction is due to fire next.

The second direction uses the advent of parallel architectures such as
Graphics Processing Units (GPU) in order to perform many realizations
of the stochastic algorithm to speed-up the overal runtime
(\cite{Komarov2012,Zhou2011,Gillespie2012,Li2008,Gillespie2013,Li2009,Tian2005,Burrage2006,Klingbeil2011}).

In every parallel implementation, each CPU core or GPU thread runs one
realization of the SSA, which means that each GPU thread is considered
as an individual unit. Each algorithm can also implement some of the
acceleration techniques developped for serial simulation, at the cost
of the complexity of the data structure.

Then, it becomes feasible to run hundreds of realizations of the same
stochastic model and they allow a scaleup that depends on the size of
the system.

However, to the best of our knowledge, the chemical systems considered
are often small systems, ranging from few species and reactions to
less than one hundred species and reactions. Recent approaches extend
these limits (see for example the hybrid approaches of Komarov et
al.~\cite{Komarov2012} that enables larger models).

It is widely accepted that it is extremely difficult to implement
efficiently the stochastic algorthim on
GPU~\cite{Klingbeil2011,Zhou2011}. And it is then anticipated that
such approaches will be limited with the growing number of large
models that systems biology generates.

For parallel implementations, the limitation comes from the fact that
each thread of the GPU had to manage a realization of a
simulation. This means that each thread had to implement a complete
algorithm. Because the memory used by each threads on one SM of the
GPU is a limited, it is possible to launch many realizations of a
stochastic simulations only for small models.

In the literature on stochastic simulation, the selection from the
propensity distribution in order to select the next reaction index $j$
is done using an inversion transformation (IT) method. 
This method uses one uniform random number $u$ and iterate in the 
cumulative distribution of the (normalized) propensities until it 
exceed $u$.
This choice is justified by the poor performance of classical
acceptance-rejection (AR) methods. However, the IT method involves a 
non-deterministic number of steps. From a parallel point of view,
running $K$ realizations in parallel with such an algorithm imposes that some
realizations have to wait for the others to finish, hence wasting resources
and impeding parallel performances.

Our purpose in this paper is to propose a new direction of research
effort by introducting a parallel acceptance-rejection algorithm to
select in parallel the $K$ next reactions to occur. This choice comes
from the following lines of evidence: At each time step of the SSA,
the propensity functions for every reaction can be considered as a
discrete distribution (called in this paper the propensity
distribution), from which one will select a reaction index $j$. This
propensity distribution is updated at each time step, but here, we
will focus only on the heart of the SSA, \textit{i.e.} the selection
of the next reaction to occur. This algorithm involves \textit{the
  same number of steps} that enables one to run $K$ realizations in
parallel without wasting hardware resources.

We show that this algorithm is very simple and that it uses
little memory space, which leaves more ressources for the other parts
(typically other optimizations) of a stochastic simulation algorithm.

We propose an implementation of a GPU hardware, but our algorithm can
be transposed to other parallel architectures. Therefore we want to
stay general and pedagogical and we do not embed our GPU/AR algorithm
procedure in a complete stochastic algorithm code.

Our GPU/AR algorithm addresses the heart the SSA and we envision that
many stochastic modelers would prefer to adapt our algorithm to their
optimizations and data structures rather than consider a new
implementation that might not fit their needs.

We hope that the simplicity of our GPU-AR algorithm will help
stochastic modelers to implement new simulation codes that could
address the new challenge of the stochastic simulations of large models.

\begin{methods}

\section{Methods}

\subsection{Graphics Processing Units}

Graphics Processing Units (GPUs) are a set of cores on a
graphical card that allow one to off-load multiple instances of the
same computations applied to large data sets,
referred to as a data-parallel computing model.

The basic execution unit on a GPU is a thread that synchronously
executes the same set of instructions, called a kernel, on different
cores of a single multiprocessor on different data pieces indexed by
the thread ID. Threads are grouped into blocks.  These blocks are
assigned to run on streaming multiprocessors, each of which is
composed of a programmer-defined number of threads.

GPU's can be seen as massively parallel many-core co-processors that
are capable of TFLOPS.

The NVIDIA Corporation introduced the Compute Unified Device Architecture
(CUDA)~\cite{Farber2011,Sanders2010} which enables developers to write programs for
the NVIDIA GPU using a minimally extended version of the C language.

One important point is that not all algorithms, often developped for
sequential architectures, are well-adapted to the GPU massive parallel
nature. More importantly, developers should develop new algorithms in
order to capture this massive parallel nature. This paper belong to
this effort.

\subsection{The Stochastic Simulation Algorithm}

Briefly, the system is described by $N$ molecular species ${ E_1,
  \cdots, E_N}$, represented by the dynamical state vector $X(t) = (
X_1(t), \cdots, X_N(t) )$ where $X_i(t)$ is the number of molecules of
species $E_i$ in the system at time $t$.  $M$ chemical reactions ${
  R_1, \cdots, R_M}$ describe how these $N$ species are in interaction
(See (\cite{Li2008}) for a general presentation).

Each reaction $R_j$ is characterized by a propensity function $a_j$
(with $\alpha_0 = \sum_{j=1}^M \alpha_j$) and a change state vector
$v_j = { v_{1j}, \cdots, v_{Nj}}$ (the propensity function is updated
at each time step of the simulation.)

At a given time step $t$ of the simulation, the propensity value $a_j$
allows one to compute the probability, given the state of the system
$X(t)$, that the reaction $R_j$ will occur in the next time interval
$[t, t+\tau]$. $v_{ij}$ is the change in the number of species $E_i$
  due to the $R_j$ reaction.

More precisely, two uniform random numbers $u_1$ and $u_2$ (from the
uniform distribution on $[0,1[$) are produced.  $\tau$ is given by
$\tau = { 1 \over a_0} \ln ( { 1 \over u_1 } )$ and the index $j$ of
the next reaction to occur is such that : $\sum_{j'=1}^j \alpha_{j^{'}} >
r_2. \alpha_0$ (\cite{Li2008}).
This method is known as the inversion transform method.

Then the system is updated using $v_j$, 
the next time value is set to $t \leftarrow t + \tau$.
A new iteration can be done, until the time $t$ reaches the end of the simulation time.

\subsection{A GPU acceptance-rejection algorithm (GPU/AR) for SSA}

Recall that our objective is to launch $K$ realizations of the SSA in
parallel, and more precisely to allows us to select $K$ next reactions
of occur, {\it i.e.} $K$ values of $j$ independently.

On order to exploit the massive parallelism of the GPU and design our
new algorithm, we propose to switch the selection of the next reaction
to occur from an inversion transformation (IT) method to an
acceptance-rejection (AR) method dedicated to $K$ parallel
realizations.

The AR method for SSA, in sequential programmation paradigm, could be
summarized as follows: choose randomly a reaction index $j$ ($1 \le j
\le M$) \textbf{and} generate a random number $u_{0,T}$ \textit{until}
$u_{0,T} < \alpha_j$. Then the next reaction to occur is $j$, and the time
step can be completed. The threshold $T$ is a parameter of the algorithm.

Such an algorithm could be implemented for each realization, and each
realization could be run on a separate CPU core or GPU threads, but,
it will suffer the same problem of non-deterministic steps as the IT
methods, not to mention the bad time execution performance.

In order to outpass this problem, we reason as follows: we split the
classical AR algorithm into two steps: an election step and a
selection step. Figure~\ref{Fig:schema} provides an illustration of
our algorithm.

\begin{figure}[!tpb]
\centerline{\includegraphics[width=9cm]{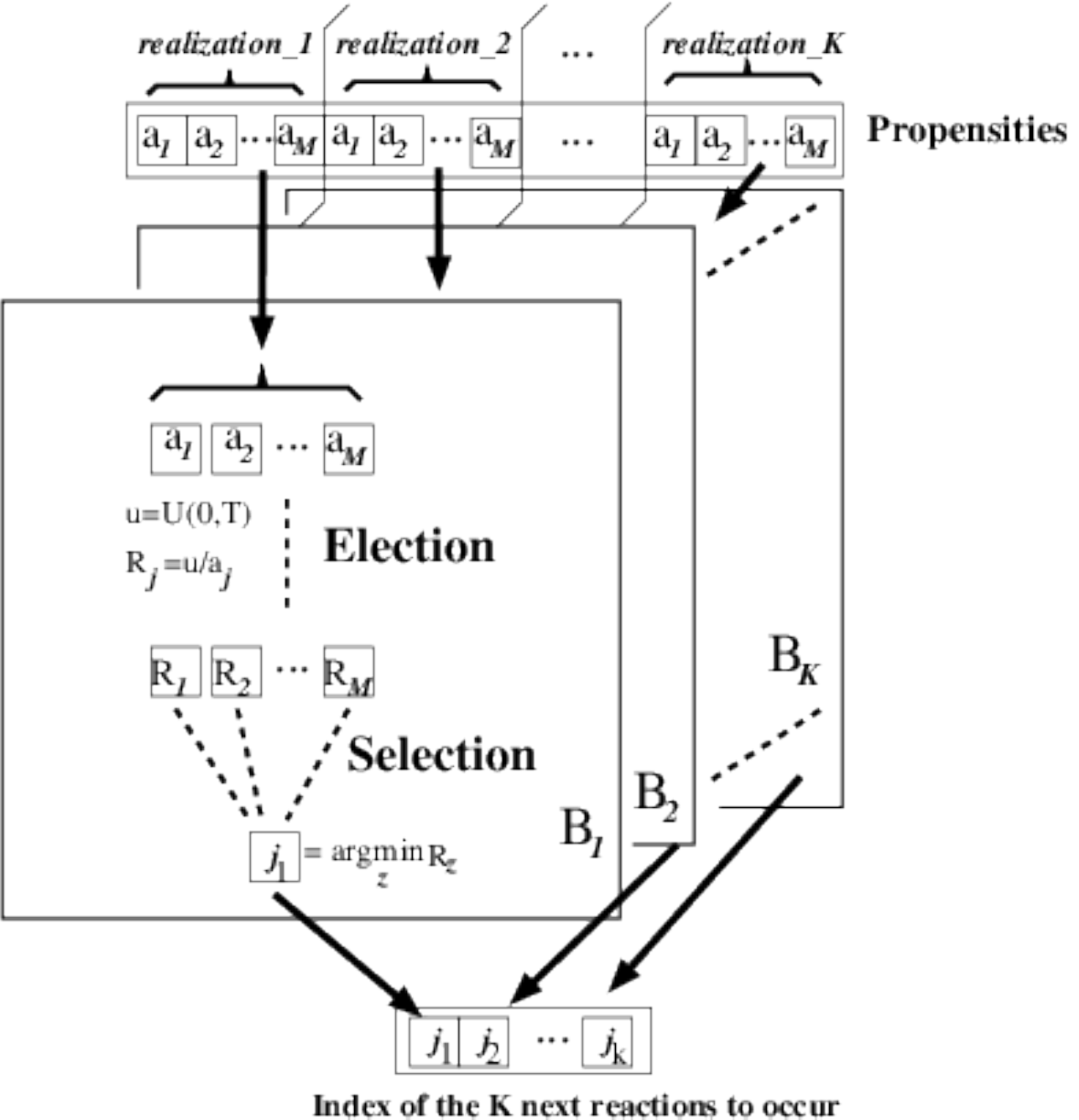}}
\caption{Diagram of the GPU acceptance-rejection algorithm.}\label{Fig:schema}
\end{figure}

In the {\bf election} step, for each reaction $j$ we test its
eligibility to occur (see below for more details) and compute a rating
value $R^k_j$ ($k$ represent the realization).

Then, in the {\bf selection} step, we choose one reaction amongst
all the eligible ones, {\it i.e.} the reaction index $j$ with the
lowest rate value $R^k_j$ for a fixed $k$.

From a parallel algorithm point of view, these two steps are
independent. Moreover, the election step is highly parallel because
each reaction $j$ can test (and rate) its eligibility independantly from other
reactions and the selection step
is done using a classical reduction method that is well suited to
parallel architectures. Several versions exist for GPU cards
with different performances (see the NVIDIA Developer Zone, 
http://developer.download.nvidia.com/assets/cuda/files/reduction.pdf).

Moreover, these two steps are also independent for each realization.

These two steps (election and selection) require $M$ threads on a
GPU. Therefore one can implement $K$ parallel instances using $K$
blocks on the GPU. This allows us to select $K$ reaction indexes $j$
at each time iteration. If $M$ is greater than the number of threads
per block, one can use several blocks ($M / maxThreadsPerBlock$) for
the election step and adapt the reduction step, but the general
philosophy of our algorithm remains unchanged.

\paragraph{Election step}

Let's denote the propensities distribution of the realization $k$ 
by $D^k$, and for the realization $k$ and for the reaction $j$ ($1 \le j \le M$)
the propensity value is denoted by $D^k_j$ 

Each reaction index $j$ tests its eligibility as follow: we generate a
uniform random number $u_{0,T}$ in the interval $[0;T[$, where $T$ is
a threshold.

If $u_{0,T} < D^k_j$ then the reaction $j$ is eligible for the
selection step and rated with the value $R^k_j = u_{0,T}/D^k_j$ else
it is not eligible (in practice every reaction is rated according
to the previous formula with the implicit rule that $R^k_j \ge 1$
means not eligible, that is a rejection).

The rate $R^k_j$ is equal to the random number $u_{0,T}$ normalized by $D^k_j$
in order to avoid bias during the selection step between reaction
indexes associated with low or high propensity values.

In classical CPU acceptance-rejection algorithms, the threshold $T$ is
very important for accuracy and rejection rate. Here, our experience
is that the choice of the threshold is not a sensitive parameter, 
and a threshold $T = \max_j [ { \alpha_j \over \alpha_0 } ]$ yields
very good results as seen in the results section.

\paragraph{Selection step}

For each realization $k$ of the SSA, 
we choose the reaction index $j$ with the lowest $R^k_j$ 
($j = \arg \min_z R^k_z$).

In order to process $K$ reductions in parallel and track the
$K$ indices $j$, we developped a modified reduction algorithm
(see~\ref{results:GPUAlgo} for an example of a GPU implementation).

These $K$ indexes $j$ represent the indices of the $K$ next reactions
to occur, one for each realization of the SSA, at the given time
iteration. Each reaction is used to compute update
the system according to $v_j$ and to compute $\tau$.

Our algorithm only uses the propensities, whatever their order, and
does not interfere with their computation and/or update, neither with
any other optimization the might by required. In this spirit, our
algorithm is not meant to replace any pre-existant optimization but to
help and add an other level of optimization in accordance to every
existing method.

\subsection{Assessing the qualities of GPU/AR algorithm}

In order to assess the quality of our algorithm, three conditions are
required. 

First, the probabilitiy for a reaction to occur should
ressemble its propensity distribution, up to a normalization
factor. We evaluated this property by computing a MSE (Mean
Squared Error) between a given propensity distribtion and the
observed probability of selection for each reaction of the model.
We used 3 simulated Gaussian propensity distributions and one propensity distribution
used for the model of iron homeostatis (see below).

Second, the rejection rate should be equal to zero as often as possible. In the
contrary, this means that some realizations of SSA would have to do a
new selection step, which will impact the parallel performances.

Third, the algorithm should be launched in a great number of parallel
intances, and moreover, it should be scalable, which means that its
performances should not be degrade when one uses a
more powerfull GPU card, or wants to increase the number of reactions
to address larger systems.

\subsection{Test and validation of the algorithm}

In order to test the quality of our algorithm, we reason that if we
consider $K$ realizations of a stochastic simulation with the
\textit{same} propensities distribution, refered to as $D$, then the
$K$ output values after one iteration step will give an estimate of
the initial propensities distribution, here refered to as $O$.  A more
accurate estimation can be obtained if we run several iterations.

We then compute the Mean Squared Error (MSE) of the
normalized distributions: $MSE = {1 \over M} \sum_{j=1}^{M} ( D_j -
O_j)^2$. This MSE will assess the precision of our algorithm.

We can also compute the rate of rejection, {\it i.e.} the number of
iterations for which the test step failed for the $M$ reactions, that
is $R^k_j > 1, \forall j$.

We generated $K$ identical propensities distributions of size $M$
(\textit{i.e.} $M$ reactions) that mimic a discrete Gaussian
$\mathcal{N}(0,1)$ using the R software~\cite{R2013}. These propensity
distributions are far from the one that arizes from stochastic
simulations, but it will facilitate the evaluation of the precision of
the algorithm.

Propensities distributions of size $M=64, 256$ and $1204$ are
generated as follow: a sequence of values $x_j \in [ -5 ; 5]$, with
$x_j = -5 + j \times { 10 \over {M-1} }$ is generated. We computed its
probability on a Gaussian distribution $\mathcal{N}(0,1)$
and multiplied it by a factor of $1e5$ in order to avoid
very small float values. Then, we drop the value $x_j$ and consider
only the index $j$, which represents our reaction index $j$.

We also use a discrete propensities distribution with $M=1024$
reactions that we developped and used in our stochastic simulations of
the iron homeostasis in Yeast~\cite{Achcar2011}.


\subsection{Materials}

Programs were written in C, using the GNU C compiler gcc version
4.4.6 on a 64-bits operating system runing Linux with 2.6.32
kernel, Intel Xeon E5-2650 (2.00GHz) with 64Go RAM

For GPU uniform random generator, we used the GPU uniform random
generator from Curand (Source: https://developer.nvidia.com/cuRAND) and the
GPU uniform random generator proposed by Michal Januszewski and
Marcin Kostur~\cite{Januszewski2010}. Only results with this latter
GPU uniform random generator will be reported.

The GPU implementation has been made in C/CUDA using the Cuda
compilation tools release 5.0 (Source:
http://www.nvidia.com/object/tesla-servers.html).

Tests and benchmarks were made on 4 different video cards : NVidia
Tesla M2075 (Fermi GPU, Peak single precision floating point
performance: 1030 Gflops, Memory bandwidth (ECC off): 150 GB/sec,
Memory size (GDDR5): 6 GBytes, Compute capability: 2.0, CUDA cores:
448), M2090 (Fermi GPU, Peak single precision floating point
performance: 1331 Gflops, Memory bandwidth (ECC off): 177 GB/sec,
Memory size (GDDR5): 6 GBytes, Compute capability: 2.0, CUDA cores:
512), Kepler K10 (2 Kepler GK104s, Peak single precision floating
point performance: 2288 Gflops per GPU, Memory bandwidth (ECC off):
160 GB/sec per GPU, Memory size (GDDR5): 4 GBytes per GPU, Compute
capability: 3.0, CUDA cores: 1536 per GPU) and Kepler K20 (1 Kepler GK110,
Peak single precision floating point performance: 3.52 Tflops, Memory
bandwidth (ECC off): 208 GB/sec, Memory size (GDDR5): 5 GBytes,
Compute capability: 3.5, CUDA cores: 2496).


\section{Results}

\subsection{Implementation}
\label{results:GPUAlgo}

For each reaction index $j$ ($0 \leq j < M$
in the C/CUDA implementation), our algorithm can be
written in pseudo-code as follows. This pseudo-code uses one
dimensional grid and blocks geometry on the GPU in order to stay
simple. This 1D-geometry limits the maximum of realizations to 65536.
The propensity value of reaction $j$ for realization $k$ is
accessed as a linear array: $D[ k*M + j]$.

The array $R$ stores the rating values for each reaction $j$ and each
realization $k$, and has a size of $M \times K$ floats.

\vskip 0.5cm
\begin{verbatim}
DEF electionStep( D, R, T ):
   D : propensities distribution [ array of float, K x M ]
   R : rate of each candidate    [ array of float, K x M ]
   T : threshold for acceptance  [ float ]

   tid : blockIdx.x*blockDim.x + threadIdx.x

   u0T = uniform_random_number_in [ 0 ; T ]
   % Default : reaction not selected (set to Max)
   R[tid] = T * 10

   if u0T < D[tid] and D[tid] <> 0:
      % the reaction is selected and rated
      R[tid] = u0T/D[tid]
\end{verbatim}
\vskip 0.5cm

For each realization $k$, this step point out which reaction could
occur and its associated rate value $R^k_j$.  The next step is to
select one reaction among all reactions that are eligible to occur. In
this work, we select the reaction with the lowest rate value, that is
$j_{next} = \arg \min_z R^k_z$.

We developped a modified reduction algorithm in order to handle $K$
reductions in parallel and tracked the index during the search for the
minimum value.  The array $I$ stores the $K$ indices of the reactions
that will occur.

\vskip 0.5cm
\begin{verbatim}
DEF SelectionStep( R, I, M )
   # R: rate array [ array of float, K x M ]
   # I: output array [ array of integer, M ]
   # M: number of reactions

   tid = blockIdx.x * blockDim.x + threadIdx.x;

   # Arrays in shared memory
   sR : rate array for propensity distribution [size M]
   sI : index value of sR                      [size M]

   # store each indice using shared memory
   sI[tid] = threadIdx.x;
   sR[tid] = R[tid];
   __syncthreads();

   # do reduction on shared memory
   # and track also the indices
   ...

   # return only the indice of the min
   if threadIdx.x == 0:
      if sR[0] < 1.0 :
         I[ blockIdx.x ] = sI[0]
      else:
         I[ blockIdx.x ]  = blockDim.x+1;
\end{verbatim}
\vskip 0.5cm

If the $R^k_j > 1$ for all $j$ that means that the election failed and
we encoutered a rejection and a value of $M+1$ is returned that
enables further post-processing.

\subsection{Quality of the selection of the reaction indexes}

For each of the four test propensities distributions, we select a
total of 10.000.000 reaction indexes using different numbers of
parallel realizations on the GPU (parameter $K$) and using different
thresholds $T_w = w \times \max_j \alpha_j$.  Two thresholds are
reported here, $T_1$ and $T_2$.

The values of $K$ are set to $100, 1000, 10000, 50000$ and $62500$.

Table~\ref{Tab:01} reports the greatest (worst situation) MSE observed
for 10 runs for each distributions and for each pair of parameters
($K, T$) for the M2090 GPU Card.

\begin{table}[!t]
\processtable{MSE between a given propensities distributions and the
  observed probability of occurence on the 10.000.000 reaction indexes
  $j$ with $K$ realizations in parallel, and for two thresholds $T_w =
  w \times \max_j alpha_j$ for $1 \le j \le M$,
  where $M$ is the number of reaction in the system.  The rejectance
  rate is always equal to 0\label{Tab:01}. GPU Cards: M2090}
{\begin{tabular}{cc|rr}\toprule
K       & N     & MSE/$T_1$ & MSE/$T_2$ \\\midrule
\hline
100	& 64	& 1.565E-07 &	1.591E-07\\
1000	& 64	& 1.612E-07 &	1.581E-07\\
10000	& 64	& 1.626E-07 &	1.614E-07\\
50000	& 64	& 1.602E-07 &	1.609E-07\\
62500	& 64	& 1.645E-07 &	1.612E-07\\
100	& 256	& 1.147E-09 &	1.171E-09\\
1000	& 256	& 1.001E-09 &	1.124E-09\\
10000	& 256	& 1.152E-09 &	1.068E-09\\
50000	& 256	& 1.107E-09 &	1.162E-09\\
62500	& 256	& 1.192E-09 &	1.155E-09\\
100	& 1024	& 1.175E-10 &	1.198E-10\\
1000	& 1024	& 1.146E-10 &	1.170E-10\\
10000	& 1024	& 1.133E-10 &	1.075E-10\\
50000	& 1024	& 1.149E-10 &	1.166E-10\\
62500	& 1024	& 1.152E-10 &	1.136E-10\\
100	& 1024	& 2.685E-07 &	2.709E-07\\
1000	& 1024	& 2.680E-07 &	2.686E-07\\
10000	& 1024	& 2.709E-07 &	2.693E-07\\
50000	& 1024	& 2.701E-07 &	2.719E-07\\
62500	& 1024	& 2.693E-07 &	2.699E-07\\\botrule
\end{tabular}}{}
\end{table}

All MSE are between $10^{-10}$ (gaussian discrete distribution with
$M=256$ and $M=1024$) and $10^{-7}$ (gaussian discrete distribution
with $M=64$ and the iron model).

No effect of the number of parallel realizations $K$ is observed,
which is expected. More importantly, no effect of the threshold has
been observed.

This means that at each step, a reaction is selected if at least one
reaction has a non nul propensity.

Our results shows that our GPU-AR algorithm performs very well because
the probability that a given reaction occur is very close, up to a
normalization factor, to its propensity.

Due to our choice a one-dimensional geometry of the grid of blocks we
launch $K=62500$ realizations of our GPU-AR algorithm in parallel
using $M=1024$ reactions. However to the best of our knowledge, this
number is larger than any previous published set of stochastic
simulations on a GPU card.

Using a two-dimensional grids of blocks, we ware able to launch one
million of realizations (data not shown).

\subsection{The rejection rate is always equal to zero}

The rejection rate is equal to zero in all our experiments: such a
result is the consequence of the parallel acceptance step. The reason
is that we used a threshold $T$ equal to the maximal value of
$\alpha_j$. This means that, unless every reaction had a null
propensity value, there is always at least one selected candidate
after the selection step. This value is selected in one step because
of the parallel implementation, instead of a random number of steps
in a classical CPU implementation.

Recall that this does not introduce a bias because the rating value of
each eligible reaction index $j$ is normalized before the selection
step.

\subsection{Performance and scalability of the algorithm}

Figure~\ref{Fig:02} reports the mean time execution (in msec) for the
same sets of experiments as in the previous section for the M2075 GPU
Card for numbers of parallel realizations on the GPU, and for
different model sizes (M=64, M=256 and M=1024). Detail of
performances for other GPU cards are shown in the panel for K=62500
parallel realizations on the GPU and for different size of the model.

\begin{figure}[!tpb]
\centerline{\includegraphics[width=9cm]{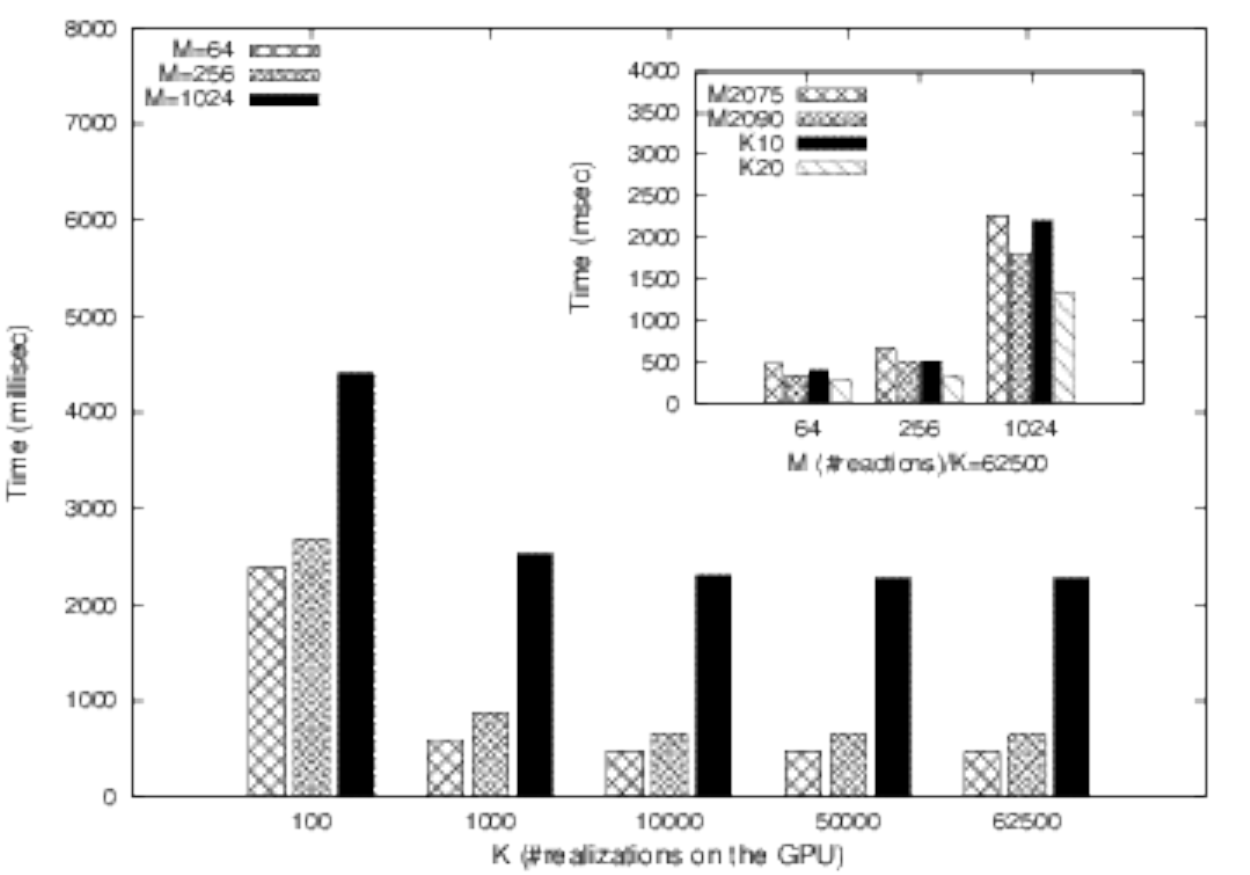}}
\caption{Main panel: mean execution time (millisec) of the GPU/AR
  algorithm on the M2075 GPU Card (averaged on 10 runs) for different
  numbers of parallel realizations. Three numbers of reactions are
  reported (M=64, M=256 and M=1024). Panel: mean
  execution time (millisec) for 4 GPU cards ( M2075, M2090,
  K10 and K20) for K=62500 parallel realizations.}\label{Fig:02}
\end{figure}

From these experiments, we can draw several conclusions. First, when
the number of reactions $M$ increases, the execution time
increases. However, when the number $K$ of realizations increases
the execution time decreases which means that the GPU is more efficiently used
and that our algorithm exploits more efficiently the GPU architecture.

These experiments also show that our GPU/AR algorithm is a scalable
algorithm: the execution time decreases when using more powerfull GPU
cards.  For example, for $M=1024$ the execution time is 2277.18 msec
on a M2075 GPU Card, and decreases to 1326.78 msec (41.73\%) on a K20
GPU Cards, This means that our GPU/AR algorithm will immediately
benefits from more efficient GPU cards. This also ensures its perenity
for future GPU cards.

\end{methods}

\section{Discussion}

Stochastic simulation in the new age of systems biology is facing the
challenge of dealing with large systems characterized by an increasing
number of reactions, and the need to run more and more realizations of
this model.

It is well-known that the heart of the SSA algorithm, namely the
selection of the next reaction to occur, is the most computationally
costly part of the algorithm. To the best of our knowledge, every
implemtentations are based on the inversion transform method, which
needed a non deterministic number of steps obviously impacting a
parallel implementation in order to tackle many realizations in
parallel.  Most of the SSA implementations on GPUs implement each
realizations as a kernel, with the restrictions imposes to each
instance of a kernel by the GPU hardware.

Here we propose a new direction of efforts that fully parallelize all
the realizations of the SSA on the GPU.  We propose to use an
acceptance-rejection algorithm to select the next reaction to occur
for $K$ parallel realizations.

We show that our algorithm: i) is well adapted to a GPU
implementation, ii) is very accurate (MSE values between a given
propensity distribution and the distribution of the observed generated
propensities), iii) allows to select the next reaction to occur for
65536 realizations using a one-dimensional geomery of grid of blocks,
iv) is scalable, meaning that our GPU-AR algorithm will benefit from
more powerfull GPU cards to be released in the future.

Our algorithm uses GPU memory for the uniform random generator (one
array of size $K \times M$ in this paper), the propensities values for
each of the $K$ realizations (one array of size $K \times M$) and an
integer array of size $K$ in order to store, at each kernel call, the
index of the next reaction to occur. The rest of the computation used
the shared memory (two arrays of size $M$, the number of reactions).

So our algorithm leaves some global GPU memory free for other
optimizations.

For the purpose of being pedagogical we present a version that uses a
maximum of $M=1024$ reactions of the model under consideration, and
uses a 1D-dimensional geometry for the blocks of $M$ threads.
However, if $M$ is larger it is possible to adapt the algorithm in
order to use $w = { M \over NumberOfThreadsPerBlock}$ blocks for the
{\it election} step and modify the reduction used for the {\it
  selection} step.  Moreover, with a 2D-dimensional grid of blocks it
is possible to extend the number of realizations in parallel.
Depending on the GPU card and its global memory, we succeed in running
one millions of realizations in parallel (data not shown).

In order to address the widest audience, we present only the heart of
the SSA that we propose to change, without embedding it in a complete
implementation that could obscure the main idea.  We hope that it will
allows its implementation in current SSA algorithm with the benefit of
using already implemented optimizations or data structures.

\section*{Acknowledgement}
We would like to thank the NVIDIA Technology Center (PSG Cluster) for 
providing access to K10 and K20 GPU Cards.

\bibliographystyle{unsrt}

\end{document}